\documentstyle[12pt,epsf]{article}
\begin{document}
\hfill{KRL MAP-272}

\hfill{JLAB-THY-00-39}

\hfill{RBRC-141}


\vspace{0.5cm}

\begin{center}
{\Large {\bf Parity-Violating Electron-Deuteron Scattering}}
\vspace{1.5cm}

{\bf L.\ Diaconescu} 

\vspace{0.25cm}
{\it Kellogg Radiation Laboratory,
California Institute of Technology, Pasadena, CA 91125}

\vspace{0.25in}
{\bf R.\ Schiavilla} 

\vspace{0.25cm}
{\it Jefferson Lab, Newport News, Virginia 23606}  \\
and \\
{\it Department of Physics, Old Dominion University,
Norfolk, Virginia 23529}  \\

\vspace{0.25in}
{\bf U.\ van Kolck}

\vspace{0.25cm}
{\it Department of Physics, University of Arizona, Tucson, AZ  85721} \\
and \\ 
{\it RIKEN-BNL Research Center, 
Brookhaven National Laboratory, Upton, NY 11973} \\
and \\
{\it Kellogg Radiation Laboratory, 
California Institute of Technology, Pasadena, CA 91125} 
\vspace{0.50in}

\begin{abstract}
The longitudinal asymmetry due to
$Z^0$ exchange is calculated
in quasi-elastic electron-deuteron scattering
at momentum transfers $|Q^2| \simeq 0.1$ GeV$^2$
relevant for the SAMPLE experiment.
The deuteron and $pn$ scattering-state wave functions
are obtained from solutions of a Schr\"odinger equation
with the Argonne $v_{18}$ potential.
Electromagnetic and weak neutral one- and two-nucleon
currents are included in the calculation.  The two-nucleon
currents of pion range are shown to be identical to those
derived in Chiral Perturbation Theory.
The results indicate that two-body contributions to the asymmetry
are small ($\simeq$ 0.2\%) around the quasi-elastic peak, but become
relatively more significant ($\simeq$ 3\%) in the high-energy wing of
the quasi-elastic peak.
\end{abstract}

\vspace{0.30in}
pacs: 13.10.+q, 21.45.+v, 25.30.Fj

\end{center}

\pagebreak

\section{Introduction}
\label{sec:intro}

With results from the SAMPLE experiment available 
\cite{sample-pro}, 
there is now great
interest in parity-violating electron-deuteron scattering. 
The
SAMPLE Collaboration measures at Bates the longitudinal
asymmetry in polarized quasi-elastic electron scattering
on the proton \cite{sample-pro} and deuteron \cite{sample-deu}. 
The asymmetries are sensitive to the nucleon's strange magnetic
and axial form factors.  If other effects are under control, the two
measurements can be used to determine the values of the form factors
at $|Q^2|=0.1$ GeV$^2$.  Neglecting two-nucleon current effects,
preliminary results are in disagreement with theoretical
estimates, in particular for the axial contribution \cite{bob}.
The purpose of the work presented here is to compute a theoretical value 
for the deuteron asymmetry including two-nucleon currents.

Neutral charge and current one-nucleon operators are well known
(see, for example, Ref. \cite{ying}).  The contribution of such
operators to the asymmetry was computed at various momentum
transfers in Ref. \cite{hadji}.  Several theoretical issues were
studied in detail.  For the kinematical region relevant to SAMPLE,
final-state interactions were found to be important.
It was also found that (except for very low momentum transfers)
the asymmetry in the vicinity of the quasi-elastic peak
is fairly independent of the choice of two-nucleon ($N$$N$) potential.

Two-nucleon charge and current operators have also been studied
to some extent.  Heavy-meson exchange contributions were considered
in Ref. \cite{horror}.  They were shown to be unimportant in a
calculation of the asymmetry that neglected final-state interactions.
In Ref. \cite{hwang}, an impulse approximation modified to incorporate
gauge invariance was employed.  The effects of parity-violating
$N$$N$ interactions on the deuteron wave function were found to be small.
Pion-exchange currents were included in the computation of the asymmetry,
but only in electromagnetic sector. 

In the present work we extend these earlier calculations.
The leading currents are calculated in Chiral Perturbation
Theory (ChPT) \cite{meissner,NNEFT} supplemented by a successful
phenomenological model \cite{carlson}.  The one-nucleon currents
considered here include phenomenological form factors and 
have the same form as those in Ref. \cite{ying}.  We also consider
pion-exchange currents in both electromagnetic and weak 
contributions, as well as effects from heavier mesons,
evaluated using the Riska prescription \cite{Ris89}.
The asymmetry is calculated with deuteron and final-state
wave functions obtained from a realistic potential,
the Argonne $v_{18}$ model \cite{Wir95}.

We present results for the kinematical region of interest to SAMPLE.
(Our calculation can be repeated at other momentum transfers, such as
those of the planned SAMPLE-Lite experiment \cite{tito} and JLab's G0
experiment \cite{G0}.)  We verify that when two-nucleon effects are
turned off, our results are in agreement with those of Ref. \cite{hadji}.
We study the effects of two-body currents on the asymmetry, both near
the quasi-elastic peak where one-body processes should dominate, and
away from the quasi-elastic peak where these two-body currents could
be important.  We are able to confirm that most of the two-nucleon
contributions to the asymmetry are due to currents of pion range.
Our results show that near the quasi-elastic peak two-body currents 
give a small contribution to the asymmetry.  Away from
the peak, they become more important, and can increase
the magnitude of the asymmetry by as much as 3\%.  We find that the 
contribution to the asymmetry associated with the electromagnetic-axial
current interference response function is about 20\%.  The overall
effect of two-nucleon currents on the data of the SAMPLE experiment
is indeed small but not negligible, and is being incorporated
in the data analysis \cite{bob}.

A summary of the relevant formulas for the calculation of the asymmetry 
is given in Sec. \ref{sec:asy}.  Section \ref{sec:emop} contains
the one- and two-body electromagnetic currents, and Sec. \ref{sec:weop}
contains the one- and two-body neutral weak currents.  The connection
between our phenomenological approach and ChPT can be found in Sec.
\ref{sec:wechi}, while the numerical computation of the asymmetry is
described in Sec. \ref{sec:calc}.  Results and conclusions are presented
in Sec. \ref{sec:res}. 
\section{The Parity-Violating Asymmetry}
\label{sec:asy}

\begin{figure}[t]
\begin{center}
\epsffile{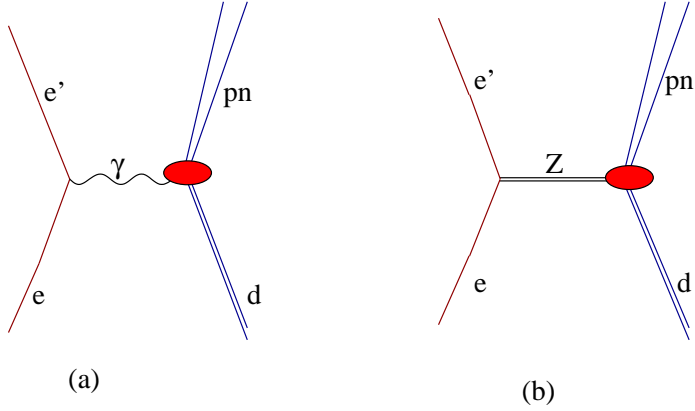}
\vskip .25cm
\caption{(a) Single photon- and (b) $Z^0$-exchange diagrams
in parity-violating quasi-elastic electron-deuteron scattering.
The blobs denote the nuclear currents.}
\label{fig:ampl}
\end{center}
\end{figure}

Parity-violating electron-nucleus scattering results from interference
of amplitudes associated with photon and $Z^0$ exchanges, shown
in Fig. \ref{fig:ampl}. 
The initial and final electron (nucleus)
four-momenta are labelled by $k^\mu$ and $k^{\prime \mu}$ ($P^\mu$
and $P^{\prime \mu}$), respectively, while the four-momentum transfer
$Q^\mu$ is defined as $Q^\mu \equiv k^\mu-k^{\prime \mu}
\equiv (\omega,{\bf q})$.  The amplitudes for the processes in
Fig.~\ref{fig:ampl} are then given by~\cite{Mus92}
\begin{equation}
M = -\frac{4\pi \alpha}{Q^2}(M^\gamma + M^Z) \ ,
\end{equation}
\begin{equation}
M^\gamma = \overline{u}^{\, \prime} \gamma^{\sigma} u\, 
            j^\gamma_{\sigma,fi} \ ,
\end{equation}
\begin{equation}
M^Z =\frac{1}{4 \pi\sqrt{2}} \frac{G_\mu Q^2}{\alpha}
 \overline{u}^{\, \prime} \gamma^{\sigma}
( g_V^{(e)}+g_A^{(e)} \gamma_5) u\, j^Z_{\sigma,fi} \ ,
\end{equation}
where $\alpha$ and $G_\mu$ are the fine-structure constant and
Fermi constant for muon decay, respectively, 
$g_V^{(e)}=-1+4\, {\rm sin}^2\theta_W$ and 
$g_A^{(e)}=1$ are
the Standard Model values for the neutral-current couplings
to the electron given in terms of the Weinberg angle $\theta_W$,
$u$ and $u^\prime$ are the
initial and final electron spinors, and $j^{\gamma,\sigma}_{fi}$
and $j^{Z,\sigma}_{fi}$ denote matrix elements of the electromagnetic
and weak neutral currents, i.e.

\begin{equation}
j^{\gamma,\sigma}_{fi} \equiv \langle f\vert j^{\gamma,\sigma}(0) \vert 
i\rangle
\equiv ( \rho^\gamma_{fi}({\bf q}), {\bf j}^\gamma_{fi}({\bf q})) \ ,
\end{equation}
and similarly for $j^{Z,\sigma}_{fi}$.  Here $\vert i\rangle$ and
$\vert f\rangle$ are the initial and final nuclear states.  Note
that in the amplitude $M^Z$ the $Q^2$ dependence of the $Z^0$ propagator
has been ignored, since here we restricted ourselves
to $|Q^2| \ll m_Z^2$.

The parity-violating asymmetry in the quasi-elastic regime is given by 

\begin{equation}
A = \Bigg ( \frac{d\sigma^+}{d\Omega d\omega}
           -\frac{d\sigma^-}{d\Omega d\omega} \Bigg )
 \Bigg / \Bigg ( \frac{d\sigma^+}{d\Omega d\omega}
           +\frac{d\sigma^-}{d\Omega d\omega} \Bigg ) \ ,
\end{equation}
where $d\sigma^h/d\Omega d\omega$ is the inclusive cross section for
scattering of an incident electron with helicity $h=\pm 1$.  It is easily
seen that, to leading order,

\begin{equation}
A \propto \frac{\Re\, ( M^\gamma M^{Z *})}{\mid M^\gamma \mid^2} \ .
\end{equation}
Standard manipulations then lead to the following expression for the
asymmetry in the extreme relativistic limit for the electron~\cite{Mus92}

\begin{equation}
A = \frac{1}{2\sqrt{2}}\frac{G_\mu Q^2}{\alpha} 
\frac{ g_A^{(e)} v_L R_L^{\gamma,0} +
       g_A^{(e)} v_T R_T^{\gamma,0} +
       g_V^{(e)} v_T^\prime R_T^{\gamma,5} }
      { v_L R_L^{\gamma,\gamma} +
        v_T R_T^{\gamma,\gamma} } \ ,
\label{eq:asy}
\end{equation}
where the $v$'s are defined in terms of electron kinematical
variables,

\begin{eqnarray}
v_L &=& \frac{Q^4}{q^4} \ , \\
v_T &=& {\rm tan}^2 (\theta/2) +\frac{|Q^2|}{2\, q^2} \ , \\
v_T^\prime &=& {\rm tan}(\theta/2)
 \sqrt{ {\rm tan}^2(\theta/2) +\frac{|Q^2|}{q^2} } \ ,
\end{eqnarray}
$\theta$ being the electron scattering angle in the laboratory, while
the $R$'s are the nuclear electro-weak response functions, which
depend on $q$ and $\omega$, to be defined below.  To this end,
it is first convenient to separate the weak current
$j^{Z,\sigma}$ into its vector $j^{0,\sigma}$ and axial-vector $j^{5,\sigma}$
components, and to write correspondingly 

\begin{equation}
j^{Z,\sigma}_{fi}=j^{0,\sigma}_{fi} + j^{5,\sigma}_{fi}
                 \equiv ( \rho^0_{fi}({\bf q}) , {\bf j}^0_{fi}({\bf q}) )
                       +( \rho^5_{fi}({\bf q}) , {\bf j}^5_{fi}({\bf q}) ) \ .
\end{equation}
The response functions can then be expressed as

\begin{equation}
R_L^{\gamma,{\rm a}}(q,\omega) = \overline{\sum_i} \sum_f \delta(\omega+m_i-E_f)
\, \Re \left [ \rho^\gamma_{fi}({\bf q}) 
               \rho^{{\rm a} *}_{fi}({\bf q}) \right ] \ ,\label{eq:rl}
\end{equation}
\begin{equation}
R_T^{\gamma,{\rm a}}(q,\omega) = \overline{\sum_i} \sum_f \delta(\omega+m_i-E_f)
\, \Re \left [ j^\gamma_{x,fi}({\bf q}) 
               j^{{\rm a} *}_{x,fi}({\bf q}) 
              +j^\gamma_{y,fi}({\bf q}) 
               j^{{\rm a} *}_{y,fi}({\bf q}) \right ] \ ,\label{eq:rt}
\end{equation}
\begin{equation}
R_T^{\gamma,5}(q,\omega) = \overline{\sum_i} \sum_f \delta(\omega+m_i-E_f)
\, \Im  \left [ j^\gamma_{x,fi}({\bf q}) 
               j^{5 *}_{y,fi}({\bf q}) 
              -j^\gamma_{y,fi}({\bf q}) 
               j^{5 *}_{x,fi}({\bf q}) \right ] \ , \label{eq:rtp}
\end{equation}
where $m_i$ is the mass of the target (assumed at rest in the laboratory),
$E_f$ is the energy of the final nuclear state (in general, 
a scattering state),
and in Eqs.~(\ref{eq:rl}) and (\ref{eq:rt}) the superscript ${\rm a}$ is either
$\gamma$ or $0$.  Note that there is a sum over the final states and an average
over the initial spin projection states of the target, as implied
by the notation ${\overline{\sum}}_i$.  In the expressions above for the $R$'s,
it has been assumed that the three-momentum transfer ${\bf q}$ is along
the $z$-axis, which defines the spin quantization axis for the nuclear states.

The model for the nuclear electro-weak currents is discussed in the next
sections, while the calculation of the deuteron response functions
is described in Sec.~\ref{sec:calc}.
\section{Electro-Weak Charge and Current Operators}

\subsection{Electromagnetic Operators}
\label{sec:emop}

The nuclear charge and current operators consist of one- and two-body terms
that operate on the nucleon degrees of freedom:

\begin{eqnarray}
\rho^\gamma({\bf q})&=& \sum_i \rho^{\gamma,1}_i({\bf q})
             +\sum_{i<j} \rho^{\gamma,2}_{ij}({\bf q}) \>\>, \label{eq1}\\
{\bf j}^\gamma ({\bf q})&=& \sum_i {\bf j}^{\gamma,1}_i({\bf q})
             +\sum_{i<j} {\bf j}^{\gamma,2}_{ij}({\bf q}) \label{eq2} \>\>.
\end{eqnarray}
The one-body operators $\rho^{\gamma,1}_i$ and ${\bf j}^{\gamma,1}_i$ have the
standard expressions obtained from a relativistic reduction of the
covariant single-nucleon current, and are listed below for convenience.
The charge operator is written as
\begin{equation}
\rho^{\gamma,1}_i({\bf q})= \rho^{\gamma,1}_{i,{\rm NR}}({\bf q})+
                       \rho^{\gamma,1}_{i,{\rm RC}}({\bf q}) \>\>, \label{eq6}
\end{equation}
with
\begin{equation}
\rho^{\gamma,1}_{i,{\rm NR}}({\bf q})= \epsilon_i \>
 {\rm e}^{{\rm i}{\bf q}\cdot {\bf r}_i} \label{eq7} \>\>,
\end{equation}
\begin{equation}
\rho^{\gamma,1}_{i,{\rm RC}}({\bf q})= \left
( \frac {1}{\sqrt{1+|Q^2| /4m^2} }-1\right )
\epsilon_i \> {\rm e}^{{\rm i}{\bf q}\cdot {\bf r}_i}
- {\frac {{\rm i}}{4m^2}} \left ( 2\, \mu_i-\epsilon_i \right )
{\bf q} \cdot (\mbox{\boldmath$ \sigma$}_i \times {\bf p}_i) \>
{\rm e}^{ {\rm i} {\bf q} \cdot {\bf r}_i } \>\>,
\label{eq8}
\end{equation}
where $|Q^2|=q^2-\omega^2 > 0$ is the four-momentum transfer defined earlier,
and $m$ is the nucleon mass.
The current operator is expressed as
\begin{equation}
     {\bf j}^{\gamma,1}_i({\bf q})={\frac {1} {2m}} \epsilon_i \>
   \bigl[ {\bf p}_i\>,\>{\rm e}^{{\rm i} {\bf q} \cdot {\bf r}_i} \bigr ]_+
   -{\frac {{\rm i}} {2m}} \mu_i \>
     {\bf q} \times \mbox{\boldmath$ \sigma$}_i \> {\rm e}^{{\rm i} {\bf q} 
     \cdot
   {\bf r}_i}  \label{eq9}\>\>\>,
\end{equation}
where $[ \cdots \, ,\, \cdots ]_+$ denotes the anticommutator.  The following
definitions have been introduced:

\begin{eqnarray}
\epsilon_i &\equiv& \frac {1}{2}\left [ G_E^S(|Q^2|) +
G_E^V(|Q^2|)\tau_{z,i} \right ] \>\>,\label{eq8a}\\
\mu_i &\equiv& \frac {1}{2}\left [ G_M^S(|Q^2|) +
G_M^V(|Q^2|)\tau_{z,i} \right ] \label{eq9a} \>\>,
\end{eqnarray}
and ${\bf p}$, $\mbox{\boldmath$ \sigma$}$, and $\mbox{\boldmath$ \tau$}$ are
the nucleon's momentum, Pauli spin
and isospin operators, respectively.  The two terms
proportional to $1/m^2$ in $\rho^{\gamma,1}_{i,{\rm RC}}$
are the well known Darwin-Foldy and spin-orbit relativistic
corrections~\cite{Fri73}, respectively.
The dipole parametrization is used for the isoscalar ($S$) and
isovector ($V$) combinations of the electric and magnetic nucleon form
factors (including the Galster form
for the electric neutron form factor~\cite{Gal71}).  

The most important features of the two-body parts of the
electromagnetic current operator are summarized below.  The reader
is referred to Refs.~\cite{Sch89,Sch90,Car90} for a derivation and listing
of their explicit expressions.

\subsubsection{Two-body current operators}
\label{sec:emop1}

The two-body current
operator has  ``model-independent'' and
``model-dependent'' components, in the
classification scheme of Riska~\cite{Ris89}.
The model-independent terms are obtained from the
two-nucleon interaction (in the present study
the Argonne $v_{18}$ interaction~\cite{Wir95} is employed),
and by construction satisfy current
conservation with it.  The leading operator is the
isovector ``$\pi$-like'' current obtained
from the isospin-dependent spin-spin and tensor interactions.
The latter also generate an isovector ``$\rho$-like'' current, while
additional model-independent isoscalar and isovector currents arise from the
isospin-independent and isospin-dependent central and momentum-dependent
interactions.  These currents are short-ranged and numerically
far less important than the $\pi$-like current.

The model-dependent currents are purely transverse
and therefore cannot be directly linked to the underlying two-nucleon 
interaction.
The present calculation includes the isoscalar $\rho \pi \gamma$ and
isovector $\omega \pi \gamma$ transition currents as well as the isovector
current associated with excitation of intermediate $\Delta$-isobar resonances
(for the values of the various coupling constants and cutoff masses
in the monopole form factors at the meson-baryon vertices, see 
Ref.~\cite{Viv00}).
Among the model-dependent currents, those associated with the $\Delta$-isobar
are the most important ones.  In the present calculation, these currents are
treated within the static $\Delta$ approximation.  While this is
sufficiently accurate for our purposes here, it is important to realize
that such an approach can lead to a gross overestimate of
$\Delta$ contributions in electro-weak transitions 
(see Refs.~\cite{Viv96,Sch92,Mar00}
for a discussion of this issue within the context of neutron and
proton radiative captures on deuteron and $^3$He, and the proton weak
capture on $^3$He).

Finally, it is worth pointing out that the contributions associated
with the $\rho \pi \gamma$, $\omega \pi \gamma$ and
$\Delta$-excitation mechanisms are, in the regime of low to moderate
momentum-transfer values of interest here ($q \leq 2 $ fm$^{-1}$),
typically much smaller than those due to the leading model-independent
$\pi$-like current~\cite{Mar98}.

\subsubsection{Two-body charge operators}
\label{sec:emop2}

While the main parts of the two-body currents are linked to the form of the
two-nucleon interaction through the continuity equation, the most important
two-body charge operators are model-dependent, and should be considered as
relativistic corrections.
Indeed, a consistent calculation of two-body charge effects in nuclei would
require the inclusion of relativistic effects in both the interaction models
and nuclear wave functions.
There are nevertheless rather clear indications for the relevance of two-body
charge operators from the failure of the impulse approximation in
predicting the deuteron tensor polarization
observable~\cite{Abb00}, and charge form factors of the three- and
four-nucleon systems~\cite{Mar98,Wir91}.
The model commonly used~\cite{Sch90} includes the $\pi$-, $\rho$-, and
$\omega$-meson exchange charge operators with both isoscalar and isovector
components, as well as the (isoscalar) $\rho \pi \gamma$ and (isovector)
$\omega \pi \gamma$ charge transition couplings, in addition to the
single-nucleon Darwin-Foldy and spin-orbit relativistic corrections.
The $\pi$- and $\rho$-meson exchange charge operators are constructed from the
isospin-dependent spin-spin and tensor interactions (those of the
Argonne $v_{18}$ here), using the same
prescription adopted for the corresponding current operators.

It should be emphasized, however, that for $q \le 2$ fm$^{-1}$ the
contributions due to these two-body charge operators
are very small when compared to those from the one-body operator.  

\subsection{Weak Operators}
\label{sec:weop}

In the Standard Model the vector part of the neutral
weak current is related to the isoscalar ($S$) and isovector
($V$) components of the electromagnetic current, denoted
respectively as $j^{\gamma,\sigma}_S$ and $j^{\gamma,\sigma}_V$, via

\begin{equation}
j^{0,\sigma}=-2 \,{\rm sin}^2\theta_W \, j^{\gamma,\sigma}_S
+(1-2\, {\rm sin}^2\theta_W)\, j^{\gamma,\sigma}_V \ ,
\end{equation}
and therefore the associated one- and two-body weak charge and
current operators are easily obtained from those given
in the preceding section.

The axial charge and current operators too have one- and
two-body terms. Only the axial current

\begin{equation}
{\bf j}^5 ({\bf q})= \sum_i {\bf j}^{5,1}_i({\bf q})
             +\sum_{i<j} {\bf j}^{5,2}_{ij}({\bf q}) \label{eq2a} \>\>
\end{equation}
enters in the calculation of the asymmetry.
The axial charge operator is not needed
in the present work.  The one-body axial current
is given, to lowest order in $1/m$, by

\begin{equation}
{\bf j}^{5,1}_i({\bf q})=-G_A(|Q^2|) \frac{\tau_{z,i}}{2} 
{\mbox{\boldmath$ \sigma$}}_i
\, {\rm e}^{{\rm i}{\bf q}\cdot {\bf r}_i} \ ,
\label{eq:axc1}
\end{equation}
where the nucleon axial form factor is parametrized as

\begin{equation}
G_A(|Q^2|)=\frac{g_A}{(1+|Q^2|/\Lambda_A^2)^2} \ .
\end{equation}
Here $g_A$ is the nucleon axial coupling constant, $g_A=1.2654$, and
the cutoff mass $\Lambda_A$ is taken to be
1 GeV/c$^2$, as obtained from an analysis
of pion electroproduction data~\cite{Ama79} and measurements of the
reaction $\nu_\mu\, p \rightarrow \mu^+\, n$~\cite{Kit83}. 

There are relativistic corrections to ${\bf j}^{5,1}$ as well as
two-body contributions arising from $\pi$-, $\rho$-, $\rho\pi$-exchange 
mechanisms and $\Delta$ excitation~\cite{Mar00}.
All these effects, however, are neglected
in the present study.  The reasons for doing so
are twofold: firstly, axial current contributions to the asymmetry
are small, since they are proportional to the electron neutral weak
coupling $g_V^{(e)} \simeq -0.074$ (see Eq.~(\ref{eq:asy}));
secondly, axial contributions from two-body operators
are expected to be at the $\simeq 1$ \% level of those 
due to the one-body operator in Eq.~(\ref{eq:axc1}).
For example, in the proton weak capture on
proton at keV energies~\cite{Sch98}--this process is
induced by the charge-changing axial weak
current--the $\pi$, $\rho$, $\rho\pi$, and $\Delta$ two-body
operators increase the predicted one-body cross section by 1.5\%.
Such an estimate is expected to hold up also in the quasi-elastic regime
being considered here.

\section{Connection with Chiral Perturbation Theory}
\label{sec:wechi}

The purpose of this section is to relate the operators presented 
in Sections \ref{sec:emop} 
and \ref{sec:weop} to ChPT.
We consider only up and down quarks, in which case
the QCD Lagrangian has an
approximate \(SU(2)_L \times SU(2)_R \) chiral symmetry. 
This symmetry is spontaneously
broken by the vacuum to the diagonal \( SU(2)_V\) subgroup, and 
three pseudoscalar Goldstone bosons, the pions \( \pi_a\),
appear in the spectrum.  Chiral symmetry provides important constraints on
the description of low-momentum processes involving pions.
In particular, it allows one to estimate the relative size of 
various contributions.

To accomplish this, the most general effective Lagrangian with broken  
\(SU(2)_L \times SU(2)_R \) is constructed.  This effective Lagrangian 
includes terms with arbitrary number of derivatives and powers
of the quark masses, however higher-dimension operators 
are suppressed by inverse powers of the characteristic mass scale 
of QCD, $M_{QCD}\sim 1$ GeV.
Thus pion interactions are determined
as a power series in \( (q, m_\pi)/M_{QCD}\), where $q$ is the typical
external three-momentum.
At low energies this is a small number and hence
only the lowest-order terms are considered in this paper.

We start with the \(\pi \pi\) and \(\pi N\) Lagrangians.
Details can be found, for example, in Refs. \cite{meissner,NNEFT}.
We lump the pions in a field 
$U = u^2=1 + {\rm i} \mbox{\boldmath$ \tau$} \cdot
\mbox{\boldmath$ \pi$}/f_\pi - \mbox{\boldmath$ \pi$}^2/2f_\pi^2 +...$,
where $f_\pi=93$ MeV is the pion decay constant.
We denote the nucleon field of velocity $v_\mu$ and spin $S_\mu$ by $N$.
In order to obtain the currents required for the computation 
of the asymmetry we must use the proper covariant
derivatives.  The covariant derivatives on the pion and nucleon field are 
constructed in terms of external vector and axial-vector
fields ${\cal V}_{\mu}$ and ${\cal A}_{\mu}$ in the usual manner.
They are given by
\begin{equation}
D_{\mu}U = \partial_{\mu}U-{\rm i}({\cal V}_{\mu}+{\cal A}_{\mu})U
+ {\rm i}\, U({\cal V}_{\mu} -{\cal A}_{\mu}), \\
\end{equation}
\begin{equation}
D_{\mu}N = \partial_{\mu}N + \frac{1}{2}[u^\dagger,\partial_{\mu}u]N -
\frac{\rm i}{2}\Bigg[ u^{\dagger}({\cal V}_{\mu}^{(3)}+{\cal A}_{\mu}^{(3)})u 
  +u ({\cal V}_{\mu}^{(3)}-{\cal A}_{\mu}^{(3)})u^{\dagger}\Bigg] N 
   -3{\rm i}\,{\cal V}_{\mu}^{(0)}N,
\end{equation}
where the superscripts $(0)$ and $(3)$ denote
isoscalar and isovector components.
It is convenient to construct also other quantities
that transform covariantly.  For example,
\begin{eqnarray}
a_\mu &=& {\rm i}[u^\dagger, D_\mu u]_+, \\
f_{\mu \nu}^\dagger &=& u^\dagger F_{\mu \nu}^Ru + uF_{\mu \nu}^Lu^\dagger, 
\end{eqnarray}
where 
$F_{\mu \nu}^{R,L} = \partial_{\mu}F_\nu^{R,L}-\partial_{\nu}F_\mu^{R,L}
-{\rm i} [F_\nu^{R,L},F_\mu^{R,L}]$ with
$F_{\mu}^R ={\cal V}_{\mu}+{\cal A}_{\mu}$ and 
$F_{\mu}^L ={\cal V}_{\mu}-{\cal A}_{\mu}$.

One can find the relation
between the external fields and $Z^0$ or photon by considering the 
covariant derivative on the quark fields (see, e.g., Ref. \cite{peskin}):
\begin{equation}
{\cal A}_{\mu}=\frac{g}{2 \cos \theta_W}\frac{\tau_z}{2}Z_\mu, 
\end{equation}
\begin{equation}
{\cal V}_{\mu}=\frac{g}{2 \cos \theta_W}(\frac{\tau_z}{2} 
                       - 2 \sin^2 \theta_W
                  Q_q)Z_\mu + eQ_qA_\mu,
\end{equation}
where $Q_q$ is the quark-charge matrix.

From these building blocks we can 
write the chiral Lagrangians,
\begin{equation}
{\cal L}_{\pi\pi} = \frac{f_\pi^2}{4} {\rm Tr} [D_{\mu}U^\dagger D^{\mu}U] 
                  + \ldots, \\
\end{equation}
\begin{eqnarray}
{\cal L}_{\pi N} = N^\dagger [{\rm i}\,v\cdot D+g_A S\cdot a]N+ \frac{1}{2m}
 N^\dagger\Bigg[(v\cdot D)^2-D\cdot D -
{\rm i}\,g_A [S\cdot D, v\cdot v]_{+} \Bigg] N 
\nonumber\\
-\frac{\rm i}{4m}N^\dagger [S^\mu ,S^\nu ]\Bigg[(1-\kappa_v) 
f_{\mu \nu}^\dagger
+\frac{1}{2}(\kappa_s-\kappa_v){\rm Tr}(f_{\mu \nu}^\dagger)\Bigg]N
+\ldots,
\end{eqnarray}
where $\kappa_s = -0.12$
and $\kappa_v = 3.71$, and $\ldots$ denote terms with more derivatives 
and/or powers of the pion mass.
One can also write down interactions containing four or more nucleon
fields \cite{NNEFT,ray}, 
which are important for a fully consistent description of systems 
involving two or more nucleons.
One and two-body currents can be obtained from these interactions.
\subsection{Ordering}
\label{sec:pc}
The symmetries allow an infinite number of interactions,
so an ordering scheme is necessary for predictive power.
We want to estimate the size of matrix elements of
one- or two-body currents between $N$$N$ wave functions.
These matrix elements involve: the final $N$$N$ wave function,
a two-nucleon propagator, the current operator,
another two-nucleon propagator, and the initial $N$$N$ wave function.
Each two-nucleon propagator brings a factor $m/q^2$, and each loop
a $q^3/(4\pi)^2$.  Apart from loops in the operators themselves,
matrix elements of one- and two-body operators involve, respectively,
one and two loops.

Let us first consider only strong interactions
between nucleons \cite{NNEFT}.  These start at ${\cal O}(q^0/f_\pi^2)$, 
and most structures that appear
up to ${\cal O}(q^3/f_\pi^2 M_{QCD}^3)$ \cite{ray} are present in the
Argonne $v_{18}$ potential \cite{Wir95}. 
Use of phenomenological potentials in conjunction
with currents derived in ChPT has already proven to be a
successful approach \cite{NNEFT}.

Contributions to the amplitude $M^\gamma$ start at ${\cal O}(e^2m/q^3)$ 
with the tree-level one-body charge operator.
First corrections come in tree-level one-body currents
from ${\cal O}(q/m)$ magnetic corrections in the Lagrangian.
Second corrections are of two types:
(i) one-loop corrections and ${\cal O}(q^2/M_{QCD}^2)$
interactions in one-body currents; and (ii) tree-level
two-body currents.  And so on.  Contributions to the amplitude $M^Z$
follow the same pattern, but have an extra overall factor of 
${\cal O}(q^2/M_Z^2)={\cal O}(G_\mu q^2/e^2)$.  The leading
one and two-body contributions are described in Sec.~\ref{sec:chptob}
below. 

Electromagnetic interactions between nucleons are clearly
higher-order effects, also included in the $v_{18}$ potential.
Interesting are also the contributions to the asymmetry that come not
from the exchange of a $Z^0$ between electron and deuteron,
but from $Z^0$ exchange between hadrons (and photon exchange between
electron and deuteron).  Let us denote these contributions by 
$M^{\gamma(Z)}$.  How does $M^{\gamma(Z)}$ compare to $M^Z$?

Direct $Z$ exchange between nucleons is an ${\cal O}(G_\mu)$ effect,
so suppressed by ${\cal O}(G_\mu f_\pi^2)$ compared to the strong
interaction.  On the other hand, the electron-deuteron
interaction is ${\cal O} (e^2/q^2 G_\mu)$ larger
in $M^{\gamma(Z)}$ than in $M^Z$.
As a consequence, the contribution from
$Z^0$ exchange between nucleons to
$M^{\gamma(Z)}$ is ${\cal O}(e^2 f_\pi^2/q^2)$
compared to the leading contribution to $M^Z$.
Since we are interested in $q \sim f_\pi$, this is
$\sim e^2$, a small effect.  Yet, we are here searching
for subleading contributions.
As we have seen, two-body currents in $M^Z$ are 
suppressed by ${\cal O}(q^2/M_{QCD}^2)$ compared to the leading one-body
effect.  Thus, compared to two-body currents in $M^Z$,
$M^{\gamma(Z)}$ can be ${\cal O}(e^2 f_\pi^2 M_{QCD}^2/q^4)$;
for momenta $q \sim f_\pi$, this is $\sim 1$.
There exist also potentially larger contributions to $M^{\gamma(Z)}$.
The size of the parity-violating pion-nucleon
coupling might be as large as ${\cal O}(G_\mu f_\pi^2 M_{QCD}/q)$ compared to
the parity-conserving coupling \cite{claudiojaime}.
In this case pion exchange in the one-body current (anapole
form factor), in the potential and 
in the two-body current could all produce an effect ${\cal O}(M_{QCD}/q)$
compared to two-nucleon effects in direct $Z^0$ exchange.

Our calculation involves all contributions suppressed
by ${\cal O}(q^2/M_{QCD}^2)$ (and also some even more suppressed) 
compared to leading, except for $M^{\gamma(Z)}$.
However, contributions to $M^{\gamma(Z)}$ from the parity-violating potential
were already examined in Ref. \cite{hwang}.
Also, although not included in any deuteron calculation,
the anapole form factor of the nucleon seems too small
\cite{claudiojaime} to be relevant here, and
it is reasonable to expect that two-body currents
stemming from the parity-violating pion-nucleon
coupling will be equally unimportant.
(This should be eventually confirmed in an explicit calculation.)
On the other hand, the leading two-body currents in the
amplitude $M^Z$ stemming from pion exchange are being
calculated here for the first time.
\subsection{One- and two-body currents}
\label{sec:chptob}
\begin{figure}[t]
\begin{center}
\epsffile{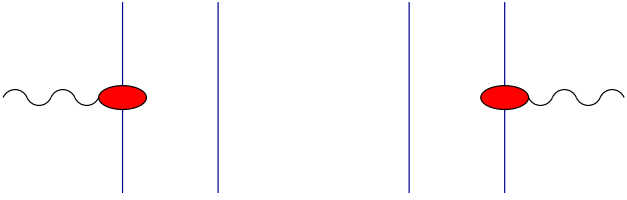}
\vskip .25cm
\caption{Electromagnetic one-body diagrams. Solid lines denote
nucleons, and the blob represents the one-body current.}
\label{f1}
\end{center}
\end{figure}

\begin{figure}[t]
\begin{center}
\vskip .75cm
\epsffile{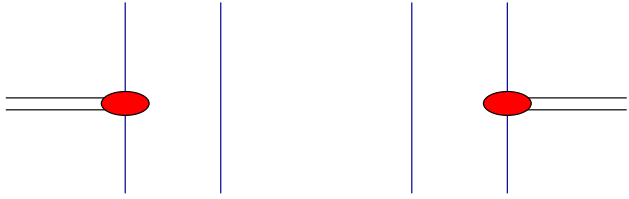}
\caption{Weak neutral one-body diagrams. Symbols as in Fig. \protect\ref{f1}.}
\label{f2}
\end{center}
\end{figure}

The one-body currents contributing 
to the processes in Fig. \ref{fig:ampl}, 
depicted in Figs. \ref{f1} and \ref{f2}, are given
to ${\cal O}(e q/M_{QCD})$ by

\begin{eqnarray}
\rho^{\rm a}({\bf r}) &=& \frac{1}{2}\sum_{i}
           (g_E^S+g_E^V\tau_{z,i})\delta({\bf r}-{\bf r}_i) \\
{\bf j}^{\rm a}({\bf r}) &=& \frac{1}{4m}\sum_{i}
           \Bigg[ (g_E^S+g_E^V\tau_{z,i})
   [\delta({\bf r}-{\bf r}_i){\bf p}_i+{\bf p}_i\delta({\bf r}-{\bf r}_i)]
\nonumber \\
    && + (g_M^S+g_M^V\tau_{z,i})
        \mbox{\boldmath$ \nabla$}\delta({\bf r}-{\bf r}_i)\times
\mbox{\boldmath$ \sigma$}_i \Bigg]\\
\rho^5({\bf r}) &=& -\frac{1}{4m}g_A\sum_{i}
[\mbox{\boldmath$ \sigma$}_i\cdot {\bf p}_i
            \delta({\bf r}-{\bf r}_i)+
        \delta({\bf r}-{\bf r}_i)\mbox{\boldmath$ \sigma$}_i\cdot {\bf p}_i]
        \tau_{z,i} \\
{\bf j}^5({\bf r}) &=& -\frac{1}{2} g_A\sum_{i}
         \delta({\bf r}-{\bf r}_i)\mbox{\boldmath$ \sigma$}_i \tau_{z,i}.
\end{eqnarray}
where ${\rm a}=\gamma$ or $0$, the
coupling constants $g_E^{(S,V)}$ and $g_M^{(S,V)}$ are 
given in Table \ref{table1}, and $g_A$ is the nucleon axial coupling constant. 
These results are in agreement with those presented in Ref. \cite{ying}.
These coupling constants acquire, in higher orders, a $Q^2$ dependence,
see Ref. \cite{ffinchpt}.  In our calculation we use a phenomenological
parametrization of the $Q^2$ dependence, as described in
Secs. \ref{sec:emop} and \ref{sec:weop}.

\begin{table}[bth]
\begin{center}
\begin{tabular}{|c|c|c|}
\hline
Form Factor & $\gamma$ & $Z$ \\
\hline
\(g_E^S \) & 1 & \(-2 \sin^2\theta_W\) \\
\(g_E^V \) & 1 & \(1-2 \sin^2\theta_W\) \\
\(g_M^S\) & \(1+\kappa_s\) & \(-(1+\kappa_s) \sin^2\theta_W\)\\
\(g_M^V\) & \(1+\kappa_v\) &
\((1+\kappa_v)(1-2 \sin^2\theta_W)\) \\
\hline
\end{tabular}
\end{center}
\caption{Coupling constants appearing in one-body currents to 
${\cal O}(e q/M_{QCD})$ in ChPT.}
\label{table1}
\end{table}

Note that ${\cal L}_{\pi\pi}$
contains a $\pi$-$Z^0$ mixing term in the form 
$Z\cdot \partial \pi^0$, which turns out to be proportional to 
$Q_{\mu}$, the four-momentum transfer.  Its contraction with the
leptonic current produces a contribution
proportional to the mass of the electron. 
In the extreme relativistic limit 
for the electron under consideration here,
this contribution can be neglected.

%
%
\begin{figure}[t]
\begin{center}
\epsffile{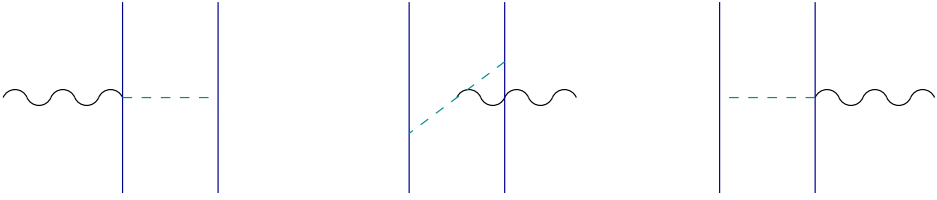}
\vskip .25cm
\caption{Electromagnetic two-body diagrams to ${\cal O}(e q^2/M_{QCD}^2)$
in ChPT.  Solid (dashed) lines denote nucleons (pions).}
\label{f3}
\end{center}
\end{figure}

\begin{figure}[t]
\begin{center}
\epsffile{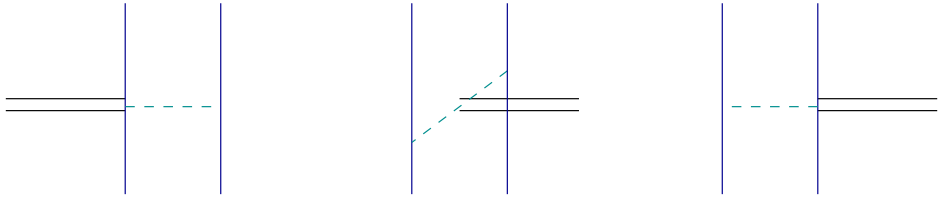}
\vskip .25cm
\caption{Weak neutral two-body diagrams to ${\cal O}(e q^2/M_{QCD}^2)$
in ChPT.  Symbols as in Fig. \protect\ref{f3}.}
\label{f4}
\end{center}
\end{figure}

The two-body contributions to the processes shown in Fig. \ref{fig:ampl}
are depicted in Figs. \ref{f3} and \ref{f4}, where again
contributions from $\pi$-$Z^0$ mixing
are neglected in the extreme relativistic limit.
To ${\cal O}(e q^2/M_{QCD}^2)$, they are given in momentum space by

\begin{eqnarray}
{\bf j}^{\rm a}({\bf k}_1,{\bf k}_2) &=& 3 {\rm i} \, g_E^V
( \mbox{\boldmath$ \tau$}_1 \times \mbox{\boldmath$ \tau$}_2 )_z
\Bigg [ v_{\pi}(k_2)\mbox{\boldmath$ \sigma$}_2 \cdot 
   {\bf k}_2 \mbox{\boldmath$ \sigma$}_1 
-v_{\pi}(k_1)\mbox{\boldmath$ \sigma$}_1 \cdot 
{\bf k}_1 \mbox{\boldmath$ \sigma$}_2 
\nonumber\\
&& \hskip2.75cm 
 - \frac{v_{\pi}(k_2)-v_{\pi}(k_1)}{{k_1}^2-{k_2}^2}
  ({\bf k}_1-{\bf k}_2)
 \mbox{\boldmath$ \sigma$}_1\cdot {\bf k}_1 
 \mbox{\boldmath$ \sigma$}_2\cdot {\bf k}_2 
 \Bigg ] \ ,
\end{eqnarray}
where ${\bf k}_i={\bf p}^\prime_i-{\bf p}_i$ with
${\bf p}_i$ (${\bf p}_i'$) denoting the initial (final) momentum 
of nucleon $i$. In the formula above,
\begin{equation}
v_{\pi}(k) =-\frac{1}{3} {\left(\frac{g_A}{4f_\pi}\right)}^2 
             \frac{1}{{m_{\pi}}^2+k^2} \ . \label{vpichpt}
\end{equation}
Note that there is no contribution to the axial current operator 
or the electromagnetic charge operator.
There is a contribution to the axial charge operator which, however,
does not enter in the asymmetry computation.
These results are in agreement with Ref. \cite{Sch89}, 
where the Fourier transform of the above expressions are
also given in detail.  

In higher order other currents appear. 
There exist shorter-range currents, which are expected
to be smaller than the ones from pion exchange with leading
order interactions.
These higher-order effects are parametrized in our calculation through
the Riska prescription
as outlined in Sections \ref{sec:emop} and \ref{sec:weop}.
In particular, we do not use Eq. (\ref{vpichpt}) for $v_{\pi}(k)$,
but the pseudoscalar component of the $v_{18}$ potential.

\section{Calculation}
\label{sec:calc}

In this section we describe the calculation of the deuteron
response functions given in Eqs.~(\ref{eq:rl})--(\ref{eq:rtp}).
The deuteron wave function is written as

\begin{equation}
\vert d,M_d\rangle=\Bigg [ \frac{u(r)}{r} {\cal Y}_{011}^{M_d}
                          +\frac{w(r)}{r} {\cal Y}_{211}^{M_d} \Bigg ]\chi_0^0 \ ,
\label{eq:deu}
\end{equation}
where the ${\cal Y}_{LSJ}^{M_J}$ are standard spin-angle functions, $\chi^T_{M_T}$
is a two-nucleon $T,M_T$ isospin state, and $u(r)$ and $w(r)$ are the S- and D-wave
radial functions. 

In the $^2$H($\vec e,e^\prime$)$pn$ reaction the final state is in the
continuum, and its wave function is written as

\begin{equation}
\vert {\bf q}; {\bf p} , SM_S , TM_T \rangle
 = {\rm e}^{ {\rm i} {\bf q} \cdot {\bf R} } \,
\psi_{{\bf p},SM_S,TM_T}^{(-)} ({\bf r}) \ ,
\end{equation}
where ${\bf r}\! =\! {\bf r}_1\! -\! {\bf r}_2$ and ${\bf R}\! =\! ({\bf r}_1 +
{\bf r}_2)/2$ are the relative and center-of-mass coordinates.  The
incoming-wave scattering-state wave function
of the two nucleons having relative momentum
${\bf p}$ and spin-isospin states $SM_S,TM_T$
is approximated as~\cite{Sch91}

\begin{eqnarray}
\psi_{{\bf p}, SM_S, TM_T}^{(-)} ({\bf r}) &\simeq &
{1 \over \sqrt{\, 2}} \Bigl[ e^{{\rm i} {\bf p} \cdot {\bf r} } -
(-1)^{S+T} e^{-{\rm i} {\bf p} \cdot {\bf r} } \Bigr] 
\chi_{M_S}^S \chi_{M_T}^T 
+{4 \pi \over \sqrt{\, 2}}
\sum_{\scriptstyle JM_J\atop\scriptstyle J\leq J_{\rm max}}
\sum_{LL'} {\rm i}^L\, \delta_{LST} \nonumber \\
&& \quad \bigl[ Z_{LSM_S}^{JM_J}(\hat {\bf p})\bigr]^* 
\Biggl[ {1 \over r} u_{L'L}^{(-)} (r; p, JST) -
\delta_{L'L}\>\> j_L (pr) \Biggr]\,
{\cal Y}_{L'SJ}^{M_J} \chi_{M_T}^T \ ,
\label{eq:pnsc}
\end{eqnarray}
where

\begin{eqnarray}
\delta_{LST} &=& 1 - (-1)^{L+S+T} \>\>, \\
Z_{LSM_S}^{JM_J} (\hat {\bf p})&=& \sum_{M_L} \langle LM_L, SM_S \vert JM_J
\rangle Y_{LM_L} (\hat {\bf p}) \>\>.
\end{eqnarray}
The $\delta_{LST}$ factor ensures the antisymmetry of the wave function,
while the Clebsch-Gordan coefficients restrict the sum over $L$ and
$L'$.  The radial functions $u_{L'L}^{(-)}$
are obtained by solving the Schr\"odinger equation
in the $JST$ channel, and behave asymptotically as

\begin{equation}
{1 \over r} u_{L'L}^{(-)} (r;p,JST)_{\,\, 
{\scriptstyle \sim\atop\scriptstyle{r\to\infty}} }\, {1 \over 2} \Biggl[
\delta_{L'L} h_L^{(1)} (pr) + (S_{L'L}^{JST})^* h_{L'}^{(2)} (pr)\Biggr] \>\>,
\end{equation}
where $S_{L'L}^{JST}$ is the $S$-matrix in the $JST$ channel and the Hankel
functions are defined as
$h_L^{(1,2)} (x)$=$ j_L (x) \pm {\rm i}\, n_L (x)$, $j_l$ and $n_L$ being
the spherical Bessel and Neumann functions, respectively.  In
the absence of interactions, $u_{L'L}^{(-)} (r; p, JST)/r \longrightarrow
\delta_{L'L}\>\> j_L (pr)$, and $\psi^{(-)}({\bf r})$ reduces to
an antisymmetric plane wave.  Interactions effects are retained in all partial waves
with $J \leq J_{\rm max}$.  In the quasi-elastic regime of interest here,
it is found that these interaction effects are negligible for
$J_{\rm max}\! >\! 7$. 

The response functions are written as (only $R_L^{\gamma,\gamma}$ is
given below for illustration)

\begin{equation}
R_L^{\gamma,\gamma} (q,\omega) =
 \sum_{S,T=0,1} R_L^{\gamma,\gamma} (q,\omega;S,T) \ ,
\end{equation}
where the contributions from the individual spin-isospin states are

\begin{equation}
R_L^{\gamma,\gamma} (q,\omega;S,T)
= {1 \over 3} \sum_{M_d M_S} \int {d {\bf p} \over
(2 \pi)^3} {1 \over 2} \vert A^\gamma_{ST} ({\bf q},{\bf p}; M_S M_d) \vert^2
\delta\Bigg(\omega+E_d - \frac{q^2}{6 \,m} - \frac{p^2} {m} \,\Bigg) \ ,
\label{eqrst} 
\end{equation}
with $A^\gamma_{ST}$ defined as

\begin{equation}
A^\gamma_{ST} ({\bf q},{\bf p}; M_S M_d) \equiv \langle {\bf q}; {\bf p},
SM_S T, M_T = 0 \vert \rho^\gamma({\bf q}) \vert d, M_d \rangle \ .
\end{equation}
Here $E_d=-2.225$ MeV is the deuteron ground-state energy,
the factor 1/2 in Eq.~(\ref{eqrst}) is included to avoid double counting, and
the states $\vert d, M_d \rangle$ and
$\vert {\bf q}; {\bf p},SM_S T, M_T = 0 \rangle$ are represented by the
wave functions in Eqs.~(\ref{eq:deu}) and (\ref{eq:pnsc}), respectively.
By integrating out the energy-conserving $\delta$-function
one finds:

\begin{equation}
R_L^{\gamma,\gamma} (q,\omega;S,T) = \frac{m\, p}{48\, \pi^2}
\sum_{M_d,M_S} \int_{-1}^{+1} d({\rm cos}\theta_{\bf p})\,
\vert A^\gamma_{ST}(q,p,{\rm cos}\theta_{\bf p};M_S M_d)\vert^2 \ ,
\end{equation}
where the magnitude of the relative momentum ${\bf p}$ is fixed by
$p=\sqrt{m(\omega+E_d)-q^2/4}$, and $\theta_{\bf p}$ is the angle
between ${\bf q}$ and ${\bf p}$.  The initial- and final-state wave
functions are written as vectors in the spin-isospin space
of the two nucleons for any given spatial configuration ${\bf r}$.
For the given ${\bf r}$ the state vector
$\rho^\gamma({\bf q}) \vert d,M_d\rangle$ is calculated
with the same methods used in quantum Monte Carlo calculations
of, for example, the charge and magnetic form factors of the
trinucleons~\cite{Mar98}.  The ${\bf r}$ and $\theta_{\bf p}$
integrations required to calculate the amplitudes
and response function are then performed by means of Gaussian
quadratures.

Finally, note that, since the deuteron is a $T=0$ state, one finds

\begin{eqnarray}
\label{eq:iso1}
R^{\gamma,0}_L(q,\omega;S,T=0) &=&
 -2\,{\rm sin}^2\theta_W \, R^{\gamma,\gamma}_L(q,\omega;S,T=0) \ , \\
\label{eq:iso2}
R^{\gamma,0}_L(q,\omega;S,T=1) &=&
 (1 -2\,{\rm sin}^2\theta_W)\, R^{\gamma,\gamma}_L(q,\omega;S,T=1) \ ,
\end{eqnarray}
with similar relations holding between the transverse response functions.
\section{Results and Conclusions}
\label{sec:res}

The asymmetry has been calculated at the kinematics relevant
to the SAMPLE experiment.  The incident electron energy was
set to \(E = 193\) MeV.  SAMPLE measures the asymmetry at four
different angles (\(\theta = 138.4^o, 145.9^o, 154.0^o, 160.4^o\)).
Different electron final energies $E^\prime$ correspond to different 
momentum transfers $Q^2$, $|Q^2| \simeq 0.1$ GeV$^2$ in the SAMPLE
experiment, which is small enough to justify the use of a non-relativistic
formalism with leading interactions obtained from ChPT.

\begin{figure}[t]
\begin{center}
\def\picsize{5in}
\epsfxsize \picsize
\epsffile{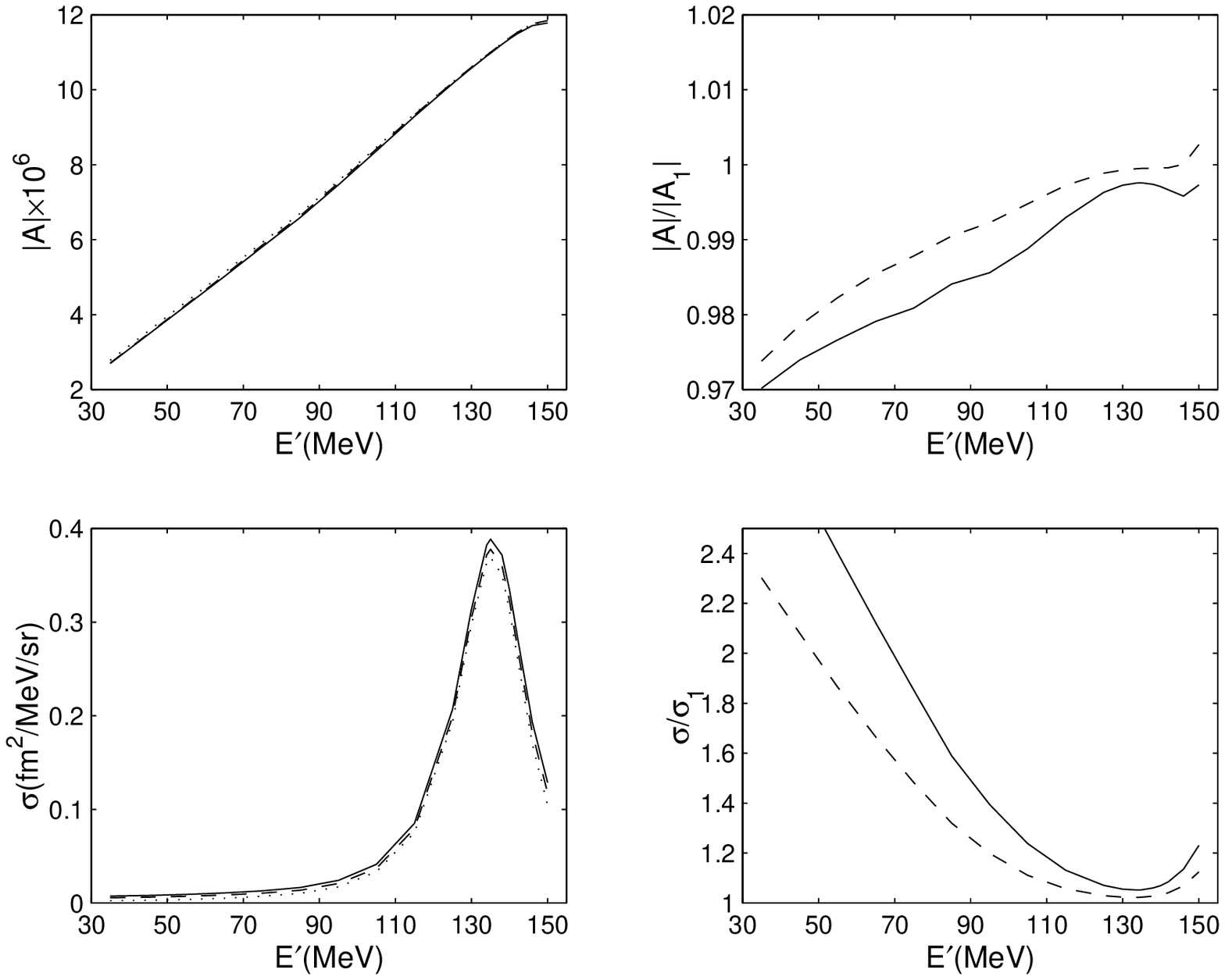}
\vskip .25cm
\caption{Results for scattering 
of an electron with incident energy $E=193$ MeV
on a deuteron at rest, as function of the electron final energy $E'$
in MeV, for a scattering angle \(\theta = 160.5^o\).
Left panels: longitudinal asymmetry $|A|$ (top) and 
cross section $\sigma$ in fm$^2$/MeV/sr  (bottom).
Shown are one-body contributions (dotted line),
one- plus two-body contributions from pion-exchange currents only 
(dashed line), and the sum of all contributions (solid line).
Right panels: ratios of one- plus two-body contributions
from pion only (dashed line) and full currents (solid line) to one-body contributions 
for the asymmetry $|A|/|A_1|$ (top) and cross section 
$\sigma/\sigma_1$ (bottom).}
\label{f10}
\end{center}
\end{figure}

\begin{figure}
\begin{center}
\def\picsize{5in}
\epsfxsize \picsize
\epsffile{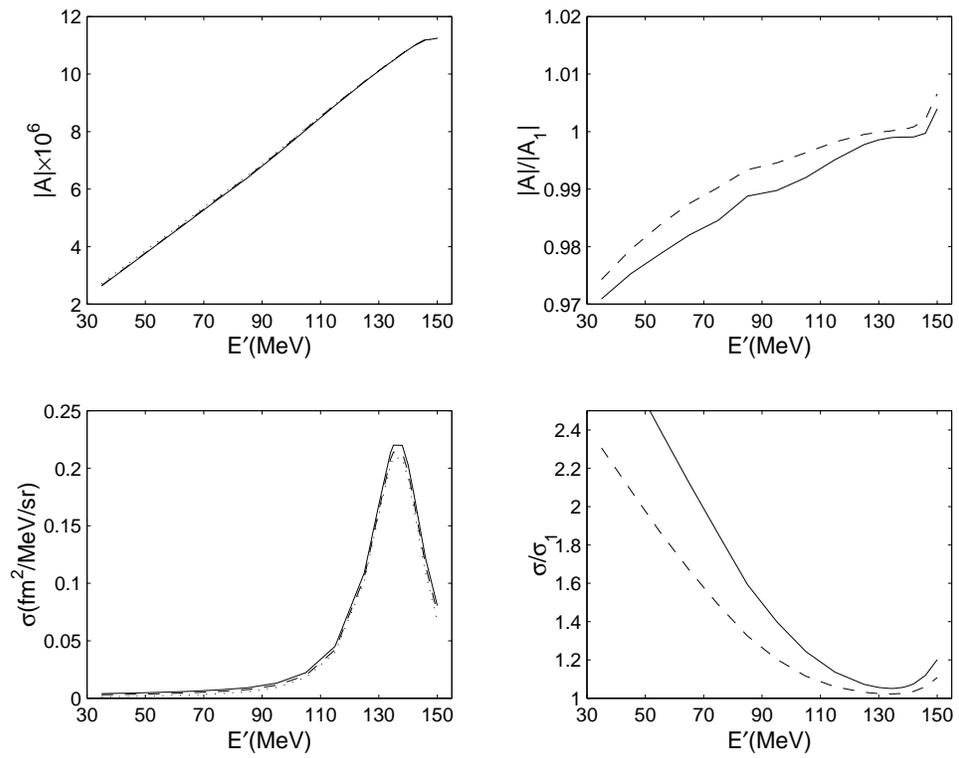}
\vskip .25cm
\caption{Same as Fig. \protect\ref{f10}, but for \(\theta = 154.0^o\).}
\label{f11}
\end{center}
\end{figure}

\begin{figure}
\begin{center}
\def\picsize{5in}
\epsfxsize \picsize
\epsffile{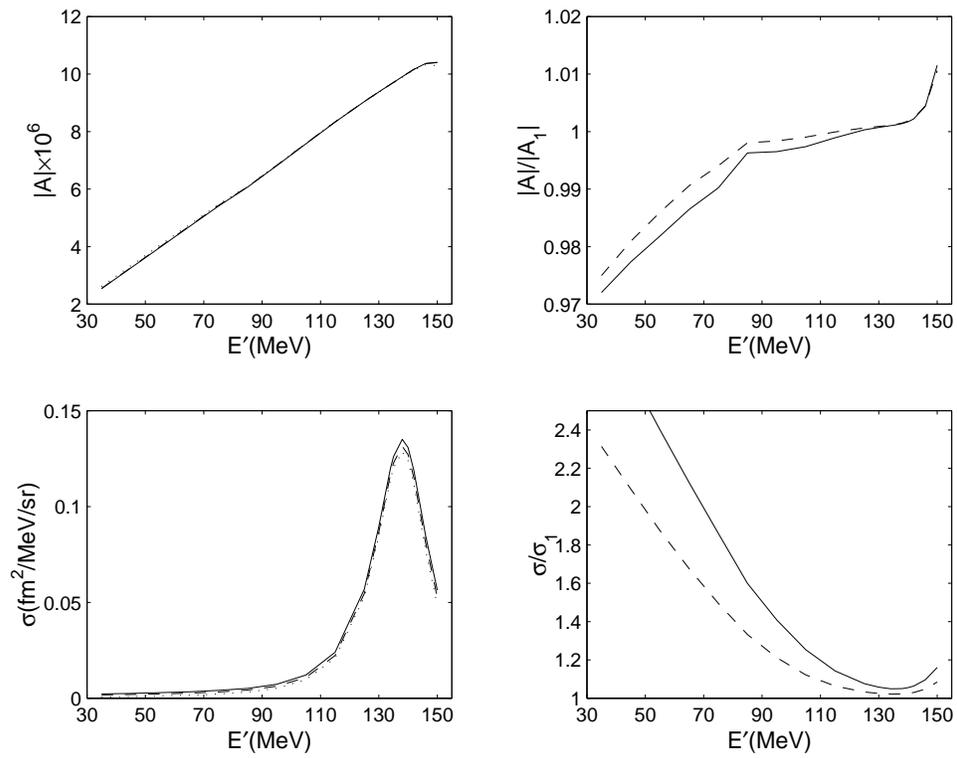}
\vskip .25cm
\caption{Same as Fig. \protect\ref{f10}, but for \(\theta =145.9^o\).}
\label{f12}
\end{center}
\end{figure}

\begin{figure}
\begin{center}
\def\picsize{5in}
\epsfxsize \picsize
\epsffile{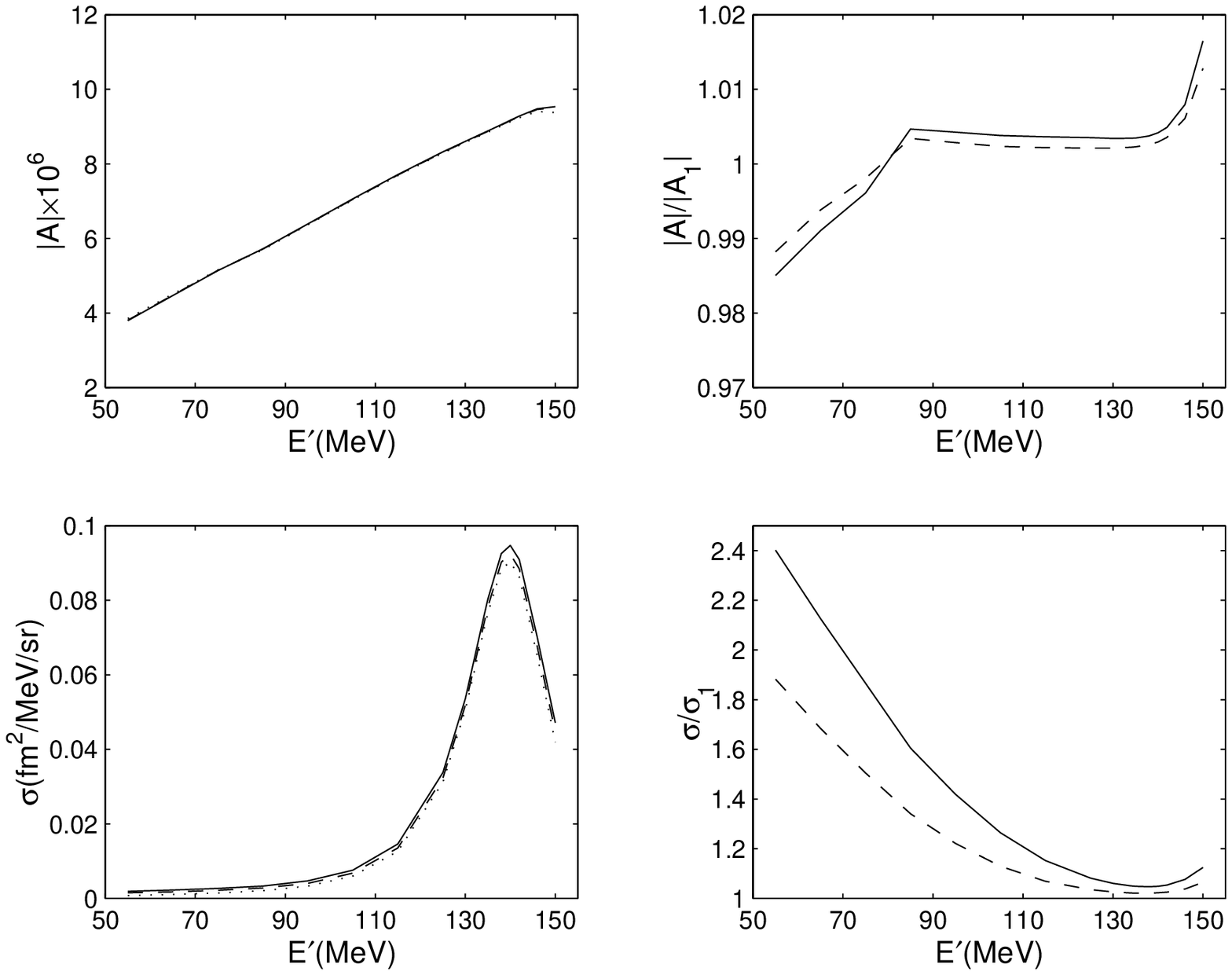}
\vskip .25cm
\caption{Same as Fig. \protect\ref{f10}, but for \(\theta =138.4^o\).}
\label{f13}
\end{center}
\end{figure}

The calculated asymmetries, as functions of the electron final energy,
are shown in Figs. \ref{f10}, \ref{f11}, \ref{f12}, and \ref{f13}, for
the four different electron scattering angles.  For each set of kinematics,
the left panels display the asymmetry and the total inclusive cross section,
with different curves representing one-body contributions, one- plus two-body
contributions from pion-exchange currents only, and the sum of all
contributions.  The ratios of one- plus two-body contributions from pion only
and full currents to one-body contributions for both asymmetries and cross
sections are displayed in the right panels.

As is apparent from the figures, the results at all angles are
qualitatively similar.  Near the quasi-elastic peak two-body 
effects in the asymmetry are negligible, less than 1\%, while
away from the quasi-elastic peak they become relatively more
important, increasing the asymmetry by at the most 3\%.
Note, however, that the two-body current contributions
are large in the inclusive cross section, indeed dominant
in the left-hand side of the quasielastic peak.  In this region
the contribution associated with the currents of pion range is more
than 50\% of the total two-body contribution.

It is interesting to examine more closely the reasons for
the relative unimportance of two-body current contributions
in the asymmetry.  At backward angles, the expression for the
asymmetry can be approximated as

\begin{equation}
\frac{A}{G_\mu Q^2/(2\sqrt{2} \alpha)}\simeq
\frac{R_T^{\gamma,0}+(-1+4\,{\rm sin}^2 \theta_W) R_T^{\gamma,5}}
{R_T^{\gamma,\gamma}} \ , 
\end{equation}
where terms proportional to the longitudinal response functions
are suppressed by the factor $v_L/v_T \leq 1/{\rm tan}^2(\theta/2)$,
a small number at the angles under consideration here (note that
the $R_L^{\gamma,{\rm a}}$ and $R_T^{\gamma,{\rm a}}$ response
functions are of the same order of magnitude).  It is useful to identify
the contributions from $T=0$ and $T=1$ $pn$ final states, and
to use Eqs.~(\ref{eq:iso1}) and (\ref{eq:iso2}), relating the
$R_T^{\gamma,0}(T=0,1)$ to $R_T^{\gamma,\gamma}(T=0,1)$.  One then
finds:

\begin{eqnarray}
\frac{A}{G_\mu Q^2/(2\sqrt{2} \alpha)}&=&
\frac{ 1-2\,{\rm sin}^2 \theta_W (1+ r^{\gamma,\gamma})
+(-1+4\,{\rm sin}^2 \theta_W) \, r^{\gamma,5}}
{1+r^{\gamma,\gamma}} \ , \\
r^{\gamma,\gamma} &=&R_T^{\gamma,\gamma}(T=0)/
R_T^{\gamma,\gamma}(T=1) \ , \\
r^{\gamma,5} &=&R_T^{\gamma,5}(T=1)/
R_T^{\gamma,\gamma}(T=1) \ .
\end{eqnarray}
Note that the $R_T^{\gamma,5}$ response function only receives contributions
from $T=1$ $pn$ final states, since the current ${\bf j}^{5,1}$ is
isovector.  The ratio $r^{\gamma,\gamma}$ is much smaller than one,
since the transverse response is predominantly isovector.  For example,
at $E^\prime=55$ MeV and $\theta=160.5^o$, $R_T^{\gamma,\gamma}(T=0)=
0.769 \times 10^{-5} (0.935 \times 10^{-5})$ MeV$^{-1}$
and $R_T^{\gamma,\gamma}(T=1)=
10.3 \times 10^{-5} (25.9 \times 10^{-5})$ MeV$^{-1}$, and hence
$r^{\gamma,\gamma}=0.0748 \> (0.0361)$ with one-body
(full) currents.  In contrast, the ratio $r^{\gamma,5}$ is of order
one; again at $E^\prime=55$ MeV and $\theta=160.5^o$,
$R_T^{\gamma,5}(T=1)=-18.3 \times 10^{-5} (-26.0 \times 10^{-5})$ MeV$^{-1}$,
and hence $r^{\gamma,5}=-1.78 \> (-1.00)$ with one-body         
(full) currents.  However, it is multiplied by the
small factor $(-1+4\,{\rm sin}^2\theta_W)=-0.074$, and so the
asymmetry turns out to be largely independent of nuclear structure
details.

\begin{figure}[t]
\begin{center}
\def\picsize{3in}
\epsfxsize \picsize
\epsffile{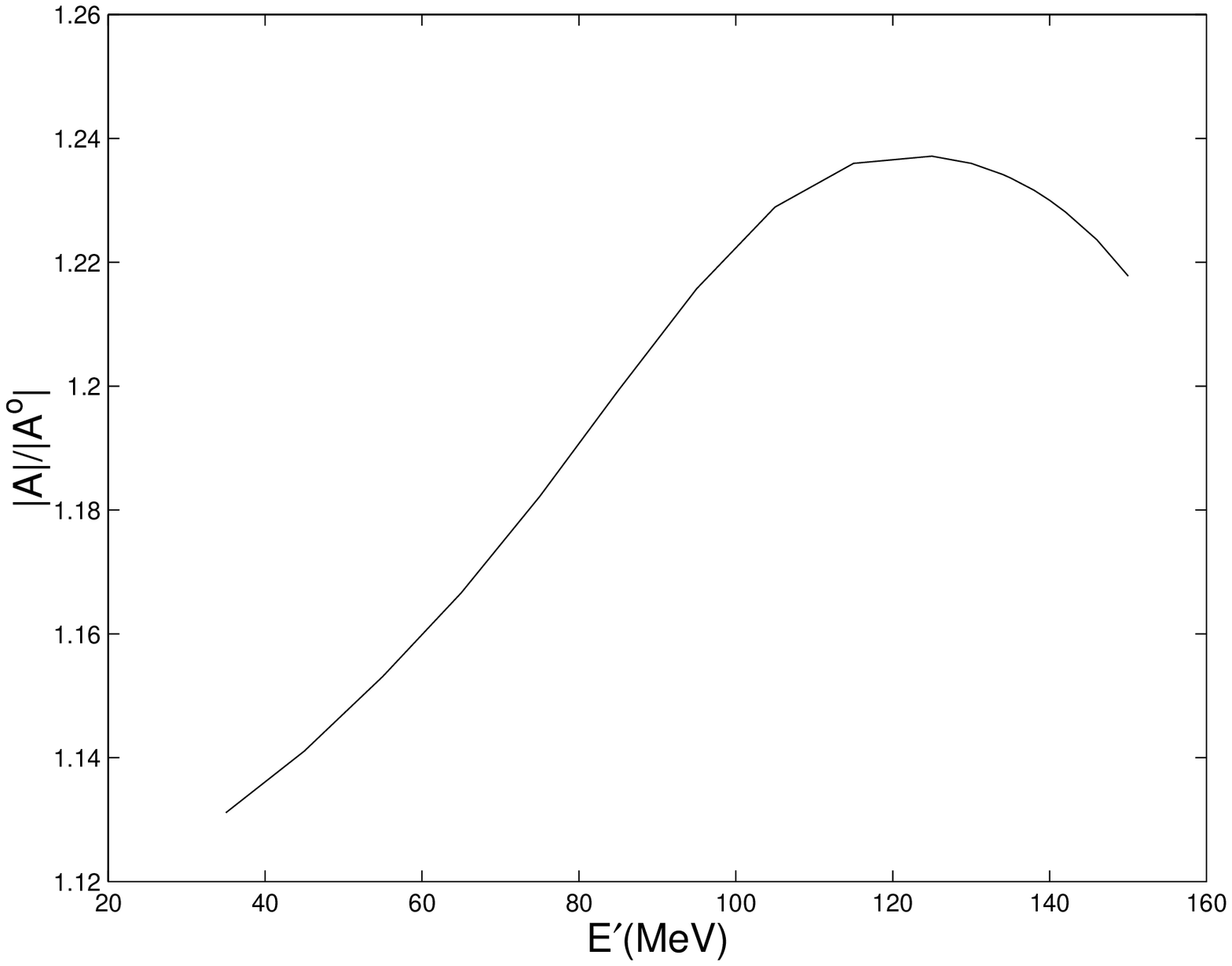}
\vskip .25cm
\caption{The ratio $|A|/|A^0|$  where  $|A|$ is the full asymmetry and
 $|A^0|$ is the asymmetry without the axial contribution 
for scattering 
of an electron with incident energy $E=193$ MeV
on a deuteron at rest, as function of the electron final energy $E'$
in MeV, for a scattering angle \(\theta = 160.5^o\).}
\label{f14}
\end{center}
\end{figure}

Finally, if $A^0$ denotes the asymmetry obtained by
ignoring the contribution of the axial current, one finds

\begin{equation}
\frac{|A|}{|A^0|} = 1+ \frac{(-1+4\,{\rm sin}^2 \theta_W) \, r^{\gamma,5}}
{1-2\,{\rm sin}^2 \theta_W (1+ r^{\gamma,\gamma})} \ .
\label{eq:ratio}
\end{equation}
The computed value for this ratio is shown in Fig. \ref{f14} for 
one of the
kinematics of the SAMPLE experiment.  The
contribution of the axial current to 
the asymmetry is of the order of 
13\% to 24\% throughout the kinematical range
considered. Note that in  Fig. \ref{f14} we have
included the small contribution from the longitudinal response.

As a last remark, we should emphasize that
the calculated transverse ($R_T^{\gamma,\gamma}$) 
and longitudinal ($R_L^{\gamma,\gamma}$) response functions
--including one- and two-body operators--
reproduce \cite{joe} existing Bates data \cite{dytman}.

Since we have performed the computations at the SAMPLE kinematics, the  above 
results may be used to account for two-body current corrections in the 
analysis of the
experimental data.  The SAMPLE experiment measures a convolution
of the asymmetry $A$ and the cross section $(d\sigma/d\Omega dE^\prime)$
over a certain range of electron final energies:
\begin{equation}
A_{total} = \frac{\int A \left(d\sigma/d\Omega dE^\prime\right) dE^\prime}
                 {\int \left(d\sigma/d\Omega dE^\prime\right) dE^\prime}.
\end{equation}
The goal of the experiment is to extract the one-body part of $A_{total}$.
To accomplish this, a model that includes
only one-body contributions is used to generate the cross section.
One can now use our results for the ratios
of total to one-body contributions in the asymmetry and 
cross section to adjust for two-body effects in the experiment.

In conclusion, we have presented a fairly complete calculation
of the asymmetry in quasi-elastic electron-deuteron scattering
arising from $Z^0$ exchange.  Since we find that, when the 
cross section is large at the quasi-elastic peak, the change
in the asymmetry due to two-body currents is negligible, we expect
that these two-body corrections will produce a modification in the
analysis of the SAMPLE experiment at the \% level,
too small to affect significantly the extraction of the strange
and axial form factors of the nucleon.
It remains to be examined whether the same holds for
effects from $Z^0$ exchange that manifest themselves
within the two-nucleon system through the parity-violating
pion-nucleon coupling.  Work along these lines is in progress.
\section*{Acknowledgements}

We are grateful to Bob McKeown for encouragement and many useful
suggestions.  L.D. thanks the Jefferson Laboratory for hospitality.
R.S. thanks the Kellogg Radiation Laboratory staff and, in particular,
Bob McKeown for the warm hospitality extended to him
during his January visits in the past three years.
U.vK. thanks Betsy Beise and Takeyasu Ito for discussions
regarding the SAMPLE experiments.
The work of L.D. and U.vK. was supported in part by NSF
grant PHY 94-20470.
The work of R.S. was supported by DOE contract DE-AC05-84ER40150
under which the Southeastern Universities Research Association (SURA)
operates the Thomas Jefferson National Accelerator Facility.
U.vK. thanks
RIKEN, Brookhaven National Laboratory and the U.S. Department of
Energy under contract DE-AC02-98CH10886 for providing the facilities essential for
the completion of this work.
Finally, some of the calculations were made possible by grants
of computing time from the National Energy Research Supercomputer
Center in Livermore.
%
%
%

%
%
%
%
%
%
\end{document}